\documentclass[aps,prl,superscriptaddress,preprint,floatfix]{revtex4-2}
\usepackage{graphicx}
\usepackage{siunitx}
\usepackage{upgreek}
\usepackage{ulem}
\usepackage[colorlinks,citecolor=blue,urlcolor=blue,bookmarks=false,hypertexnames=true]{hyperref} 
\usepackage{color}

\begin{document}

\title{Stabilization of a nonlinear bullet coexisting with a Bose-Einstein condensate in a rapidly cooled magnonic system driven by a spin-orbit torque}%
\author{Michael Schneider}
\email[e-Mail: ]{mi\_schne@rhrk.uni-kl.de}
\affiliation{Fachbereich Physik and Landesforschungszentrum OPTIMAS, Technische Universit\"at Kaiserslautern, D-67663 Kaiserslautern, Germany}
\author{David Breitbach}
\affiliation{Fachbereich Physik and Landesforschungszentrum OPTIMAS, Technische Universit\"at Kaiserslautern, D-67663 Kaiserslautern, Germany}
\author{Rostyslav O. Serha}
\affiliation{Fachbereich Physik and Landesforschungszentrum OPTIMAS, Technische Universit\"at Kaiserslautern, D-67663 Kaiserslautern, Germany}
\author{Qi Wang}
\affiliation{Faculty of Physics, University of Vienna, A-1090 Vienna, Austria}
\author{Morteza Mohseni}
\affiliation{Fachbereich Physik and Landesforschungszentrum OPTIMAS, Technische Universit\"at Kaiserslautern, D-67663 Kaiserslautern, Germany}
\author{Alexander A. Serga}
\affiliation{Fachbereich Physik and Landesforschungszentrum OPTIMAS, Technische Universit\"at Kaiserslautern, D-67663 Kaiserslautern, Germany}
\author{Andrei N. Slavin}
\affiliation{Department of Physics, Oakland University, Rochester, MI, USA}
\author{Vasyl S. Tiberkevich}
\affiliation{Department of Physics, Oakland University, Rochester, MI, USA}
\author{Bj\"orn Heinz}
\affiliation{Fachbereich Physik and Landesforschungszentrum OPTIMAS, Technische Universit\"at Kaiserslautern, D-67663 Kaiserslautern, Germany}
\author{Thomas Br\"acher}
\affiliation{Fachbereich Physik and Landesforschungszentrum OPTIMAS, Technische Universit\"at Kaiserslautern, D-67663 Kaiserslautern, Germany}
\author{Bert L\"agel}
\affiliation{Fachbereich Physik and Landesforschungszentrum OPTIMAS, Technische Universit\"at Kaiserslautern, D-67663 Kaiserslautern, Germany}
\author{Carsten Dubs}
\affiliation{Innovent e.V. Technologieentwicklung, Jena, Germany}
\author{Sebastian Knauer}
\affiliation{Faculty of Physics, University of Vienna, A-1090 Vienna, Austria}
\author{Oleksandr V. Dobrovolskiy}
\affiliation{Faculty of Physics, University of Vienna, A-1090 Vienna, Austria}
\author{Philipp Pirro}
\affiliation{Fachbereich Physik and Landesforschungszentrum OPTIMAS, Technische Universit\"at Kaiserslautern, D-67663 Kaiserslautern, Germany}
\author{Burkard Hillebrands}
\affiliation{Fachbereich Physik and Landesforschungszentrum OPTIMAS, Technische Universit\"at Kaiserslautern, D-67663 Kaiserslautern, Germany}
\author{Andrii V. Chumak}
\affiliation{Faculty of Physics, University of Vienna, A-1090 Vienna, Austria}%

\date{May 2021}%
\date{May 2021}%

\begin{abstract}

 We have recently shown that injection of magnons into a magnetic dielectric via the spin-orbit torque (SOT) effect in the adjacent layer of a heavy metal subjected to the action of short (\SI{0.1}{\micro\second}) current pulses allows for control of a magnon Bose-Einstein Condensate (BEC). Here, the BEC was formed in the process of rapid cooling (RC), when the electric current heating the sample is abruptly terminated. In the present study, we show that the application of a longer (\SI{1.0}{\micro\second}) electric current pulse triggers the formation of a nonlinear localized magnonic bullet below the linear magnon spectrum. After pulse termination, the magnon BEC, as before, is formed at the bottom of the linear spectrum, but the nonlinear bullet continues to exist, stabilized for additional \SI{30}{\nano\second} by the same process of RC-induced magnon condensation.  Our results suggest that a stimulated condensation of excess magnons to all highly populated magnonic states occurs.
 
\end{abstract}

\maketitle

Many recent experiments in the field of magnon spintronics \cite{chumak2015} make use of the injection of magnons via the application of DC currents. The mechanism responsible for injection is the spin-transfer torque (STT) effect \cite{Slonczewski1996, berger1996}, which describes the transfer of angular momentum from a spin current to the magnonic system. Once the injection overcompensates for the magnon damping, the excitation of auto-oscillations can be observed \cite{Tsoi1998, demidov2010, madami2011}.  

Due to the viscous damping of magnons \cite{Gilbert}, the lifetime of magnons is approximately inversely proportional to their energy. Subsequently, as Slonczewski and Berger predicted \cite{Slonczewski1996,berger1996}, the low-energy magnon states are auto-oscillating due to their lowest thresholds for damping compensation. Further studies found that the increase in the magnon population results in a nonlinear frequency shift,  inducing the formation of a distinct auto-oscillating mode with a frequency far below the magnon linear spectrum \cite{PhysRevLett.95.237201}. This low-frequency mode is identified as a soliton mode, referred to as the bullet mode, and observed in various STT-driven magnon systems \cite{PhysRevLett.95.237201, Demidov2016}.

Nowadays, the spin Hall effect (SHE)  \cite{PhysRevLett.83.1834} is a commonly used mechanism to generate a spin-polarized current and, combined with STT, to inject magnons \cite{urazhdin2003, Kajiwara2010, mohseni2013}. In particular, the SHE-STT effect is often studied in yttrium iron garnet / platinum (YIG / Pt) systems \cite{Collet2016, Lauer2017, haidar2016, hamadeh2014} as the SHE in Pt efficiently converts the charge current into a spin current, and the YIG exhibits a low magnetic damping. The combination of the SHE and STT effect is also named spin-orbit torque (SOT) effect \cite{PhysRevB.87.020402}.

In parallel, macroscopic quantum states in the form of magnon Bose-Einstein condensates (BECs) are intensively investigated \cite{Demokritov2006, nakata2014, Serga2014,  tserkovnyak2016, Safranski2017, Schneider2020}.  In particular, the spontaneous coherency of the magnon BEC as an interference pattern \cite{Nowik}, supercurrents under a thermal gradient \cite{ Bozhko.2016}, and the emergence of Bogoliubov waves have been observed \cite{Bozhko2019}.

Recently, we found a new approach using the rapid cooling (RC) mechanism to trigger the BEC formation. The RC mechanism results from the fast decrease of the phonon temperature \cite{Schneider2020}, and has been experimentally realized, for instance, in micro-sized YIG/Pt bilayer structures. The application of a short DC heating pulse to the Pt layer resulted in Joule heating. Subsequently, the phononic and the magnonic systems have an increased population, being in thermal equilibrium at the end of the pulse. After pulse termination, the phonon temperature decreases rapidly, which is promoted by the microscopic structure size. This fast phonon cooling results in an overpopulation of the magnon system. Suppose the number of the so-created excess magnons is large enough. In that case, the magnon redistribution to the bottom increases the chemical potential up to a critical value, and, in turn, the formation of a magnon BEC is triggered \cite{Schneider2020}. 

We further discussed the combination of the SOT effect and the RC mechanism \cite{Schneider2021} and demonstrated that the SOT effect can control the magnon BEC formation. In particular, we focused on short pulse durations and current densities overcompensating the effective damping. The application of a $100$-ns-long DC-pulse in Ref.~\cite{Schneider2021} resulted in an overpopulation at the bottom of the magnon spectrum. Still, it was kept short enough to prevent the formation of auto-oscillations in the form of the bullet mode in the current density range of interest \cite{Schneider2021, Lauer2017, Divinskiy2019}. As a result, the magnon population in a broad frequency range at the bottom increased and, consequently, the threshold current shifted.

Here, we investigate the effect of the RC mechanism after reaching a quasi-stationary auto-oscillation regime induced by the spin injection via the SOT effect. In accordance to recent studies \cite{Lauer2017, Divinskiy2019}, we observe that the STT-induced damping overcompensation over a sufficiently long time triggers quasi-stationary auto-oscillations for the longer pulse duration used. In particular, due to the induced nonlinear frequency shift, the bullet mode is formed exhibiting frequencies below the magnon dispersion. We find that excess magnons generated by the RC effect are redistributed  after pulse termination to the previously STT-driven states. Consequently, the intensity and lifetime of the bullet mode increase.

\begin{figure}[]
	\includegraphics[]{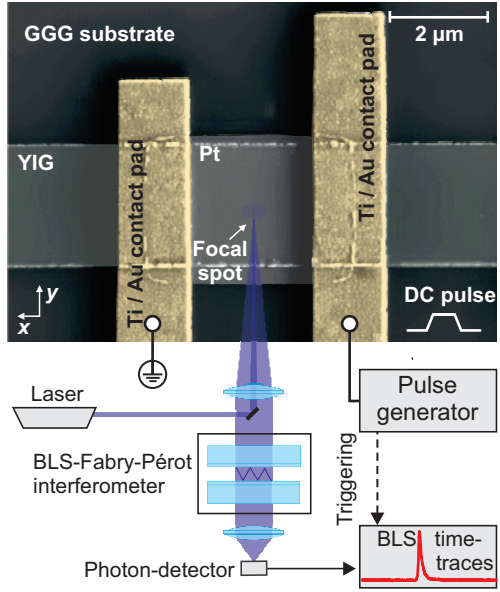}
	\caption{Colored SEM image of the structure under investigation and sketch of the experimental setup. The structure consists of a 2-$\upmu$m-broad and  34-nm-thick YIG waveguide. Ti/Au-leads contact a 3-$\upmu$m-long Platinum-heater (7~nm) on top. The spin-wave intensity under the Pt-covered region is measured using space- and time-resolved Brillouin-light-scattering spectroscopy. The Ti/Au contact pads attached to the Pt injector result in a DC current parallel to the long axis of the strip. The overlap of the contacts with the Pt-injector on both sides is 0.5~$\upmu$m, resulting in an active injection area of 2~$\upmu$m $\times$ 2~$\upmu$m.}
	\label{figure1}
\end{figure}

Figure~\ref{figure1} depicts the structure under investigation and schematically the experimental setup. The structure consists of a 2-$\upmu$m-broad waveguide, fabricated employing argon ion milling from a 34-nm-thick LPE-grown YIG film \cite{Dubs2020, Heinz.2020}. On top of the waveguide, a 3-$\upmu$m-long and 7-nm-thick Pt-layer is deposited using an RF-sputtering technique. Via electron beam evaporation, Ti/Au-leads were attached to the Pt-layer with an overlap of 0.5~$\upmu$m on each side, resulting in an active injection area of 2~$\upmu$m $\times$ 2~$\upmu$m. Two macroscopic reference pads, one made out of bare YIG and one with YIG and the Pt-layer deposited on top, were fabricated simultaneously on the same chip. Standard VNA-FMR measurements revealed 
a Gilbert damping parameter of $\alpha^\mathrm{YIG}=1.83\times10^{-4}$ and $\alpha^\mathrm{YIG/Pt}=18.1\times10^{-4}$ for the YIG and the YIG/Pt pad, respectively, and, for the latter, a corresponding spin mixing conductance of \linebreak $g^{\uparrow\downarrow}=(5.40\pm1.02)\times10^{18}\SI{}{\per\square\meter}$.

A DC pulse generator is connected to the leads, allowing for the application of DC pulses with a transition time of $\SI{1}{\nano\second}$. The application of a DC pulse results in a SHE-generated spin current, which acts on the magnetization dynamics in the YIG via the STT effect. Here, we focus on the case of a current polarity leading to magnon injection.
In addition, the DC pulse results in Joule heating of the Pt-layer. After pulse termination, the injector region cools down rapidly due to the heat diffusion promoted by the microscopic size of the heated injector and the thermal coupling to the GGG substrate \cite{Schneider2020}.  

The magnetization dynamics is investigated using time- and space-resolved Brillouin Light Scattering (BLS) spectroscopy \cite{Sebastian2015}. The light of a  laser with a wavelength of $\lambda=\SI{457}{\nano \meter}$ is focused on the YIG layer. The laser beam is guided through the transparent substrate, allowing for the investigation of magnon dynamics below the metallic structures. The inelastically scattered light carries the information about the frequency and wave-vectors of the scattered magnons and is analyzed by a 6-pass Tandem-Fabry-P\'erot-interferometer and detected by a single photon counting module. The latter connects to a time-resolution unit triggered by the pulse generator, allowing us to investigate the time evolution of the magnon population as a function of the applied DC pulse.  

To investigate the effect of the RC mechanism, which takes place after pulse termination, we first analyze the decaying STT-driven magnon system in the absence of a substantial contribution of the RC mechanism. As we have shown recently for a structure on the same chip, the STT-induced damping compensation sets in for current densities below the threshold of the RC mechanism induced condensation \cite{Schneider2021}. Although we cannot suppress the RC mechanism completely, this feature allows us to work in a supercritical STT regime without a significant RC-effect contribution. 
Moreover, the increase of the pulse duration to $\tau_\mathrm{P}= \SI{1.0}{\nano\second}$ establishing a quasi-stationary auto-oscillation regime, also decreases the RC mechanism contribution with respect to our earlier studies \cite{Schneider2020, Schneider2021}. A larger area close to the Pt-covered region is heated up, reducing heat dissipation speed and efficiency. For the particular structure and applied field, we find the threshold of the STT-driven damping compensation as $U_\mathrm{th}^\mathrm{STT}=\SI{0.87}{\volt}$ (see supplemental materials).

\begin{figure}[t]
	
		\includegraphics[]{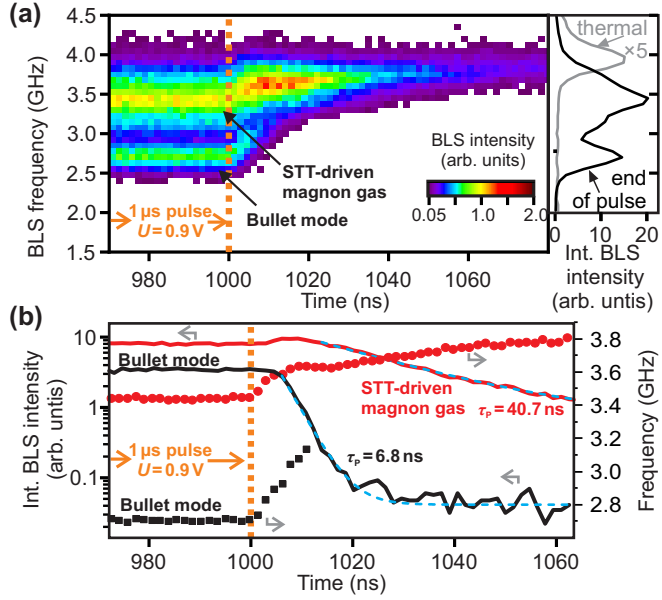}

	\caption{(a) BLS intensity color-coded as a function of time and frequency. The vertical dashed orange line shows the time the 1-$\upmu$s-long excitation current pulse is terminated.  The right panel shows the cross-section for $\SI{970}{\nano\second}<t<\SI{1000}{\nano \second} $ (black line) and, for comparison, the thermal magnon spectrum (grey line). (b) Integrated BLS intensity of the bullet mode (black line) and the bottom of the magnon gas (red line) as a function of time. Integration intervals were $\SI{2.4}{\giga\hertz}<f<\SI{3.0}{\giga\hertz}$ and $\SI{3.0}{\giga\hertz}<f<\SI{4.0}{\giga\hertz}$, respectively.  We fit the decay exponentially (blue dashed lines), revealing a significantly lower lifetime of the bullet.
		Red dots and black squares show the frequencies of both modes, same color code. The bullet mode frequency instantaneously increases after pulse termination. }
	\label{figure2}
\end{figure}
Fig.~\ref{figure2}(a) shows the BLS intensity as a function of time and frequency for a voltage of $U=\SI{0.9}{\volt}$ (see supplemental materials for whole pulse duration). It can be seen that the SOT effect increases the magnon population in two different frequency regions. The signal at $f\approx\SI{3.5}{\giga\hertz}$ corresponds to the fundamental mode: The measured frequency during the pulse is slightly below the frequency of thermal magnons, that is observed at times when no pulse is applied ($\SI{1060}{\nano\second}<t<\SI{1080}{\nano\second}$). The lower frequency results from the heating and the nonlinear shift due to the large number of magnons injected. Because of the broad line width (see corresponding cross-section), we attribute this signal to an increased population at the bottom of the magnon gas rather than the excitation of a single mode.

In addition, we observe a peak at a lower frequency of $f=\SI{2.7}{\giga \hertz}$ and with smaller line width (see cross-section). Due to the low frequency and line width, we identify this peak as the SOT-driven excitation of the bullet mode. Such a simultaneous excitation of the low-frequency region of thermal magnons and the bullet mode has been observed in similar structures before \cite{Divinskiy2019}. For the measured geometry, demagnetization fields can give rise to edge modes, which also exhibit frequencies below the dispersion of the fundamental mode. Additional space-resolved measurements and experiments with an external field aligned along the waveguide confirm the bullet mode nature of the observed peak. We discuss both in the supplemental materials. 

\begin{figure*}[t!]
		\includegraphics[width=\textwidth]{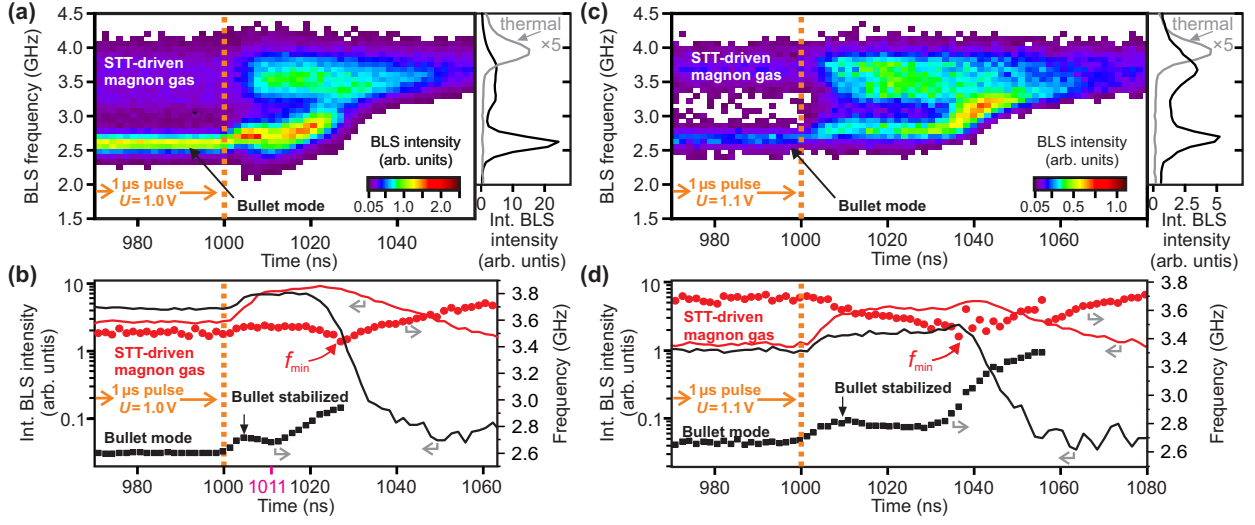}
	\caption{(a) BLS intensity, color-coded as a function of BLS frequency and time at the end and after a 1-$\upmu$s-long pulse of a voltage of $U=\SI{1.0}{\volt}$. The vertical dashed orange line indicates the end of the pulse. 
		(b) Integrated BLS intensity of the bullet mode (black line) and the bottom of the magnon gas (red line) as a function of time. Integration intervals were $\SI{2.4}{\giga\hertz}<f<\SI{3.0}{\giga\hertz}$ and $\SI{3.0}{\giga\hertz}<f<\SI{4.0}{\giga\hertz}$, respectively. Red dotes and black squares show the frequencies of both modes, same color code.
		(c,d) Same as (a,b), but for $U=\SI{1.1}{\volt}$.}
	\label{figure3}
\end{figure*}
 
We now analyze the time evolution after pulse termination, when the RC mechanism triggers an overpopulation of the whole magnon gas. Consequently, due to the magnon redistribution processes to the bottom, a deviation from a purely exponential decay is expected. As mentioned before, we expect a relatively small RC effect contribution due to the low voltage (slightly supercritical in terms of the STT-driven damping compensation) and long pulse duration.

Figure~\ref{figure2}(b) shows the integrated intensities of the bottom of the magnon gas (red line) and the bullet mode (black line) as a function of time. Further, the red dots and black squares depict the magnon gas bottom and bullet mode frequencies, respectively. Both values are obtained by fitting the sum of two Gaussians. The dashed blue lines are exponential fits.

The effect of the RC mechanism is visible as an increase in the intensity of the bottom of the magnon gas. We attribute this increase to the magnon BEC formation process, in which the lowest energy state of the linear magnon dispersion (for this case ${f_{\mathrm{{min}}} \approx\SI{3.6}{\giga \hertz}}$) attracts the excess magnons, which are initially distributed across the whole spectral range \cite{Schneider2020}. In addition, the SOT effect creates a second attractor state in the form of the bullet mode, which has an even lower frequency (here ${f_{\mathrm{{bullet}}} \approx\SI{2.7}{\giga \hertz}}$). As a result, the bullet mode intensity stays constant after pulse termination for approximately 4~ns. Albeit the bullet intensity in this short time interval being constant, we observe increasing frequencies of both modes, starting instantaneously after pulse termination, which we attribute to the bullet instability.
In this respect, we define the stability of the bullet mode as the state when it conserves its coherency and soliton nature, which manifests itself in the lower frequency of the bullet mode rather than in its intensity. Thus, if the frequency of the bullet mode increases, we refer to this process as to the bullet instability. In contrast, a constant or decreased bullet frequency reflects its stability. 
Thus, in Fig.~\ref{figure2}(b), the bullet instability in the absence of an external injection is observed as the instantaneously starting frequency increase. These observations can be understood as follows: The RC mechanism redistributes magnons to both states,  whereby the lowest energy state of the linear dispersion and the SOT-driven bullet mode of lower frequency serve as natural or artificially created attractor states, respectively.  However, the RC mechanism is ultimately too weak to stabilize the bullet mode.

Further in time, the RC-induced redistribution finishes, and we observe an exponential decay of both modes' intensities. The fits to the falling slopes reveal amplitude lifetimes of $\tau_\mathrm{P}=\SI{40.7}{\nano\second}$ for the magnon gas bottom and $\tau_\mathrm{P}=\SI{6.8}{\nano\second}$ for the bullet mode. This discrepancy shows the instability of the STT-driven bullet in the absence of an external injection again.

In the following, we discuss the situation for an increased RC effect, which we achieve via increasing the applied voltage.
Figure~\ref{figure3}(a) shows the same data as Fig.~\ref{figure2}, but now for $U=\SI{1.0}{\volt}$. It can be seen that the quasi-stationary regime establishes during the pulse and, due to the higher voltage applied, the bullet mode is now predominantly populated with respect to the bottom of the magnon gas and compared to the case shown in Fig.~\ref{figure2}.
After pulse termination (orange dashed line) and in contrast to the previous case, we now observe an increasing intensity of both modes, which directly shows that the excess magnons generated in the RC process redistribute to both  attractor states. 

To analyze the time evolution further, Fig.~\ref{figure3}(b) shows the integrated intensities in the frequency range of the bottom of the magnon dispersion (red line) and of the bullet mode (black line), the frequency of the bottom of the magnon gas (red dots) and the bullet frequency (black squares).

We first focus on the bottom of the magnon gas (red line and red dots). Initially, after pulse termination, an increasing intensity and a practically constant frequency are observed. 
For this particular case, the average frequency of the SOT-driven magnon gas (red dots at $t=\SI{1000}{\nano\second}$) is close to the final BEC frequency given by the bottom of the linear magnon dispersion (red dots at $t\approx\SI{1025}{\nano\second}$, ${f_\mathrm{min}}\approx\SI{3.4}{\giga\hertz}$). Hence, the artificially increased magnon population of the magnon gas increases the natural magnon attraction to these states. The frequency conservation is a consequence of the competition between the decreasing nonlinear shift, the cooling process and the accumulation of magnons at the lowest energy state of the magnon dispersion. Whereas the first two effects were observed in the original RC effect experiment as well \cite{Schneider2020} the varying number of dipolar magnons, which we expect to contribute most to the nonlinearity, constitutes a secondary underlying time evolution. This SOT-induced nonlinearity is the reason for the now lower final frequency of the magnon BEC with respect to the previous case of lower voltage. Finally, after approximately $t=\SI{1030}{\nano \second}$, the relaxation of the overpopulated system is visible as an increasing frequency and decreasing intensity.

We now analyze the time evolution of the bullet mode after pulse termination. Initially, the bullet mode frequency increases. This increase, analogously to the situation for $U=\SI{0.9}{\volt}$, shows the instability of the bullet. However, the increased RC effect and the stronger attraction of the excess magnons to the now more dominantly populated bullet mode are visible as an increase in the bullet mode intensity. In the following, this increased RC effect is sufficient to stabilize the bullet, and we observe a decreasing frequency once the number of redistributed magnons is high enough. Hence, two quasi-coherent states coexist in the form of the RC-driven BEC and the now stabilized bullet. The time at which the bullet stabilizes corresponds to the time-scale of the RC effect triggered redistribution processes. Following our previous findings \cite{Schneider2020}, where we observed a maximum BLS intensity around $\SI{10}{\nano\second}$ to $\SI{20}{\nano\second}$ after pulse termination, we measure the minimum bullet frequency, noting the time of highest bullet mode stability $\SI{11}{\nano\second}$ after the pulse.

Finally, for $t>\SI{1011}{\nano \second}$, we find the bullet mode frequency increases again, which shows that the RC mechanism is now too weak to stabilize the bullet. However, despite this observed instability, the bullet mode intensity is found to stay constant until $t\approx\SI{1020}{\nano\second}$. This constant intensity indicates that magnons are still redistributed to this artificially created attractor state, compensating for the damping losses. We interpret this ongoing redistribution as a thermalization process in which the magnon system transits from the auto-oscillation regime to a thermalized magnon gas. The former is given by the overpopulation of the magnon condensate and the bullet mode, and the latter quasi-equilibrated state is reached via the increase in the chemical potential, increasing the magnon number at the lowest energy state.

The situation for an even higher voltage of $U=\SI{1.1}{\volt}$ is shown in Figs.~\ref{figure3}(c,d), analogously to (a,b). Qualitatively, the same effects as for the previously discussed case are observed; we limit the discussion to the quantitative changes: With the increased voltage, the intensity of both modes at the end of the pulse decreases, as already observed in Ref.~\cite{Schneider2021}. We attribute this decrease to a decreasing spin mixing conductance with increasing temperature \cite{Uchida2014}. Furthermore, the line width of the bottom of the magnon gas decreased, which might be a consequence of the ongoing mode competition, resulting in a varying mode population for different injection rates. In particular, the frequency region below $f\approx\SI{3.5}{\giga \hertz}$ is less populated in comparison with the situation for $U=\SI{1.0}{\volt}$.

For the bullet, we find that it is stabilized significantly longer, for up to $\SI{30}{\nano\second}$ after pulse termination. We find a constant frequency of $f=\SI{2.78}{\giga\hertz}$ for $\SI{1015}{\nano\second}<t<\SI{1030}{\nano\second}$, which is higher than the stable bullet frequency at the end of the pulse of $f=\SI{2.66}{\giga\hertz}$. This higher but stable frequency of the bullet results from the lower temperature and the absence of Oersted fields after pulse termination. The bullet mode intensity is first rapidly and then still slightly increasing until $t=\SI{1040}{\nano\second}$. Consequently, we can map the time evolution of the unstable bullet even until its frequency converges with the bottom of the magnon gas.

For the magnon gas, we see a pronounced frequency decrease after the pulse, contrary to the conserved frequency in the previous case. The decrease starts at a higher frequency but ends up at approximately the same frequency of $f\approx\SI{3.4}{\giga\hertz}$ (bottom of the magnon dispersion). The now initially higher frequency results directly from the varied auto-oscillation spectral distribution at the end of the pulse: 
The now stronger populated states of the magnon gas with higher frequencies compared to the previous case increase the average frequency at the end of the pulse (compare red dots in Fig.~\ref{figure3}(c) and (d) at $t\gtrsim\SI{1000}{\nano\second}$). As a result, excess magnons are dominantly attracted to these higher-frequency states. Please note that the number of redistributed excess magnons is at least three times larger after the pulse and hence, the measured frequency is determined by the redistributed magnons. Thus, we observe that the SOT-driven increased population of these higher-frequency states attracts the excess magnons. The subsequent frequency shift is caused by the natural attraction of the excess magnons to the lowest energy state of the dispersion (BEC formation). The observed varying spectral profiles of the auto-oscillations are the topic of another project, which we will publish separately.

 Our findings show that the excess magnons generated by the RC effect and distributed across the whole spectral range are attracted to the bottom of the magnon gas and to the bullet mode serving as an artificial attractor state. As a result, during the process of rapid cooling, these two quasi-coherent states coexist. Due to the strong localization of the bullet mode \cite{PhysRevLett.95.237201}, this can be interpreted as a spatially extended condensate in the rapidly cooled region, in which the localized bullet is embedded. Due to the limited spatial resolution we cannot confirm such a spatial separation directly, even though we address the localization of the bullet mode qualitatively in the supplemental material.

In conclusion, we have shown that the RC mechanism after the termination of an applied DC pulse can stabilize the bullet mode, which is excited via the SOT effect acting during the pulse. In particular, we found an increasing intensity of the bullet mode after the pulse.
Depending on the applied voltage, the bullet mode conserves its low frequency for up to $\SI{30}{\nano\second }$ - contrary to the time evolution without the RC mechanism.

Hence, the presented results provide new evidence for the underlying physical mechanism that triggers the bullet mode formation. The RC effect might serve as an additional mechanism controlling conventional spin-torque oscillators (STOs) in the pulsed regime.

 Moreover, we report that the spectral profile of the already excited states at the time the RC mechanism sets in determines which states are populated via the subsequent RC mechanism. We interpret this feature as a stimulated condensation: The excess magnons, as bosons, want to assemble in the already highly populated  states. In our case, the SOT effect determines the initial spectral magnon population when the RC mechanism sets in.  Due to its frequency-dependence, the SOT injection populates modes in the low-frequency region. Although this low energy of the excited states might promote the condensation to these states, it seems likely from the obtained results that the RC mechanism can also trigger the redistribution to other highly populated states. Hence, we suggest that the RC mechanism can also serve as a magnon amplification mechanism in magnonic circuits.

\begin{acknowledgments} 
	This research was funded by the European Research Council within the Starting Grant No. 678309 ``MangonCircuits'' and the Advanced Grant No. 694709 ``SuperMagnonics'', by the Deutsche Forschungsgemeinschaft (DFG, German Research Foundation) within the Transregional Collaborative Research Center – TRR 173 – 268565370 “Spin+X” (projects B01 and B04) and through the Project 271741898, and by the Austrian Science Fund (FWF) within the project I 4696-N.
\end{acknowledgments}

\bibliography{apsguide4-2}
\newpage
\begin{center}
\section{Supplemental Materials}

\end{center}
\subsection{SHE-STT-induced threshold of damping compensation and spectral profiles at the end of the applied pulses}

To evaluate the threshold voltage that corresponds to a full damping compensation via the SHE-STT effect, we analyze the time evolution of the magnon intensity using the method presented in Ref. [S1]. For an effective magnon relaxation rate of  $\Gamma_\mathrm{eff}$  and a  STT-induced damping compensation, measured in terms of the parameter $\beta (j)$, the threshold of full damping compensation is reached for $\beta(j)-\Gamma_\mathrm{eff}=0$. In this case, the intensity at the beginning of the applied pulse is given by $I=I_0 \mathrm{exp}[(\beta(j)-\Gamma_\mathrm{eff})  t]=I_0 \mathrm{exp}[g(j)  t]$, where $g(j)= \beta(j)-\Gamma_\mathrm{eff}$  is the exponential growth rate of the magnon intensity. Since $\beta(j)$ is a linear function of the current, we can determine the threshold by fitting the growth rate as a function of the applied voltage, and extrapolate the fit to find the threshold condition $g(U)=\beta(U)-\Gamma_\mathrm{eff}=0$.
\begin{figure}[h]
	
		\includegraphics[width=\textwidth]{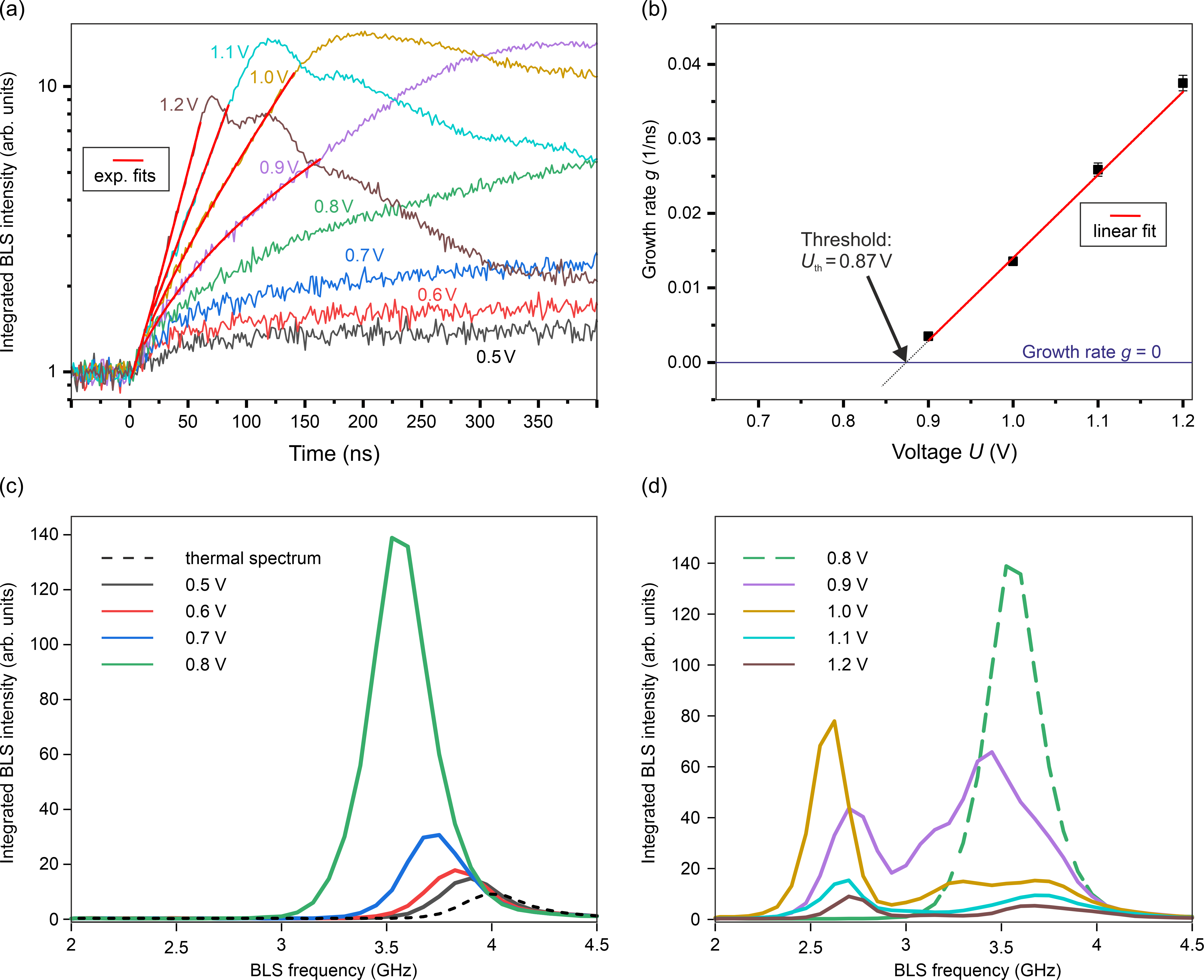}

	\caption{Integrated BLS intensity (integration range $ \SI{1}{\giga\hertz} < f < \SI{4}{\giga\hertz}$) as a function of time for different applied voltages. Red lines correspond to exponential fits, yielding the growth rates~$g$.                (b) Growth rate $g$ as a function of the applied voltage $U$. The intercept of the linear fit (red line) with        $g = 0$ (blue line) yields a threshold of $U = \SI{0.87}{\volt}$. (c) BLS spectra at the end of the pulse (integrated during the last \SI{100}{\nano\second} of the applied pulse) for voltages below the threshold. The black dashed line shows for comparison the thermal magnon spectrum. (d) BLS spectra analogous to (c), but for supercritical voltages, for comparison the dashed green line shows the spectrum for the largest applied sub-critical voltage. }
	
\end{figure}

Figure 4(a) shows integrated BLS intensity (integration range $\SI{1}{\giga\hertz} < f < \SI{4}{\giga \hertz}$) as a function of  time for different voltages applied. To determine the growth rates, we fit the rising slopes exponentially (limited to the voltages $U\geq\SI{0.8}{\volt}$, for which we observe an exponential intensity increase over time). Figure~4(b) depicts the determined growth rates $g$ as a function of the applied voltage. The intercept of the linear fit (red line) and $g(U)=0$ yields a threshold voltage of $U_\mathrm{th}=\SI{0.87}{\volt}$.
Figures~4(c,d) show the BLS spectra extracted at the end of the applied pulse when a quasi-stationary regime has been established. For comparison, the dashed line shows the thermal magnon spectrum.  For the sub-critical voltages applied [Fig.~4(c)], we observe an apparent increase of the magnon density at the bottom of the spectrum with an increasing voltage, similar to the case reported in Ref. [28]. The increased number of magnons causes a non-linear shift to lower frequencies, which increases with an increasing applied voltage. 
Figure~4(d) depicts the magnon spectra for the supercritical voltages applied, analogous to Fig.~4(c). For comparison, the dashed green line shows the spectrum for the highest applied sub-critical voltage of $U=\SI{0.8}{\volt}$.  For these supercritical voltages, the magnon spectra change substantially: In addition to the non-linear shifted magnon spectra, a sharp, distinct peak at a frequency of $f\approx\SI{2.7}{\giga\hertz}$ below the shifted spectrum appears. Due to the low frequency, we attribute this peak to the SHE-STT effect triggered generation of the bullet mode. We confirm the bullet mode nature in additional experiments for a field applied along the waveguide and by space resolved measurements, both presented in the following. In addition, the shape of the high-frequency peak, which we attribute to the non-linear shifted magnon spectrum, changes, which implies that injected magnons populate also certain distinct modes within this frequency range. The variation of the relative intensities with the applied voltage suggests that the ongoing mode competition in the supercritical regime determines the spectral magnon density increase. However, due to the limited frequency resolution of the BLS setup ($\Delta f\approx\SI{150}{\mega\hertz}$) and the broad magnon background, the nature of the excited modes remains speculative.

\subsection{BLS spectra as a function of time for the whole pulse duration}

\begin{figure}[b]
	
		\includegraphics[width=\textwidth]{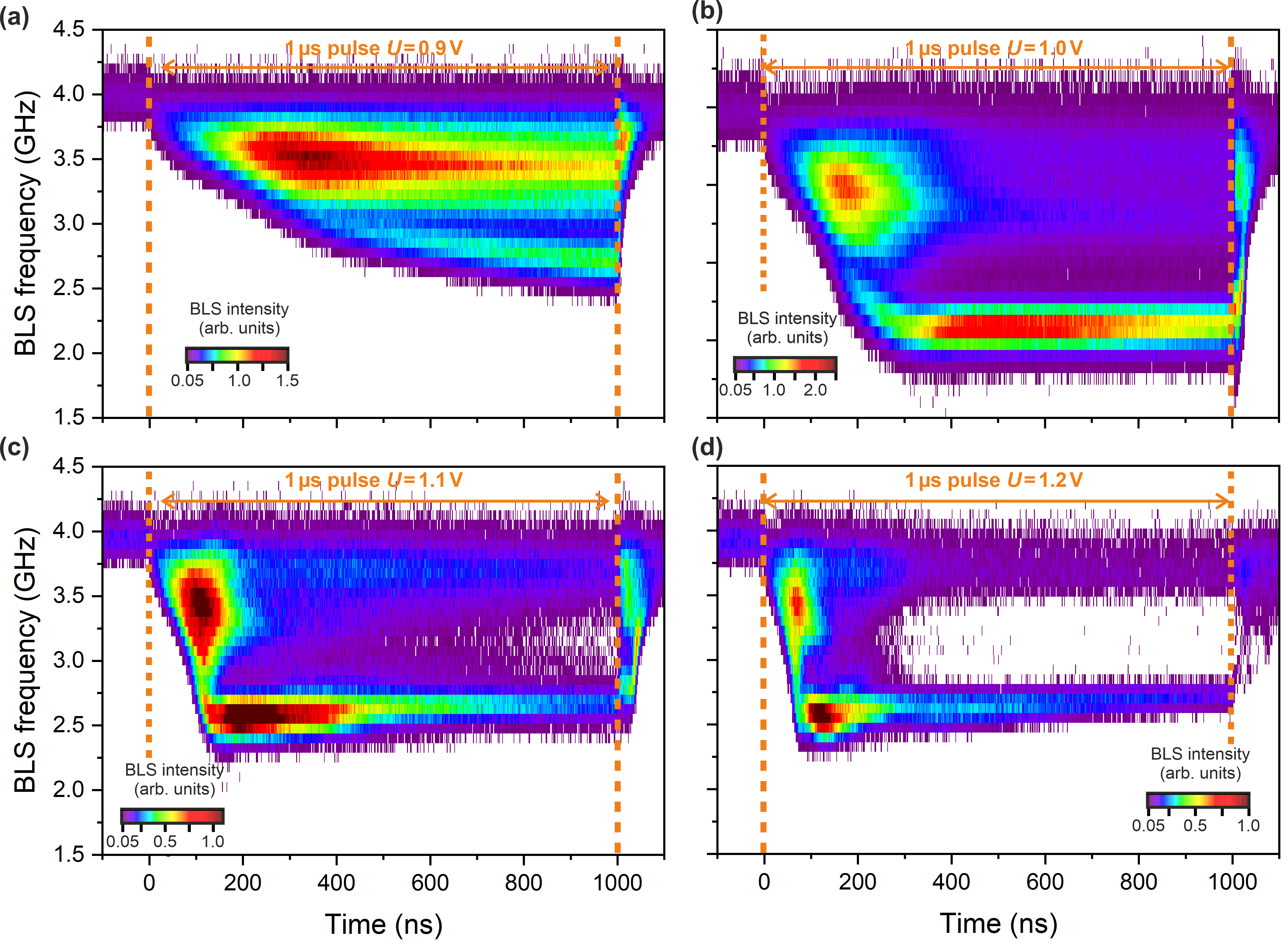}

	\caption{BLS intensity as a function of frequency and time for different applied voltages. The time a 1-$\upmu$s-long pulse is applied is indicated by the dashed orange lines. (a) $U=\SI{0.9}{\volt}$, for this lowest supercritical voltage the bullet mode ($f<\SI{3.0}{\giga\hertz}$) is formed at $t\approx\SI{0.5}{\micro\second}$. The increased magnon density at higher frequencies corresponds to the frequency-shifted thermal magnon spectrum (magnon gas). As the intensity of the bullet increases, the magnon intensity at higher frequencies decreases. (b)  $U=\SI{1.0}{\volt}$, the bullet mode is formed earlier, at  $t\approx\SI{0.2}{\micro\second}$, its intensity is increased with respect to (a), the intensity at higher frequencies decreases at the time the bullet is formed. (c ) $U=\SI{1.1}{\volt}$, a further increase of the bullet mode intensity earlier in time with respect to the situation depicted in (a,b) is observed. After the initial increase the overall intensity and the bullet intensity decrease again during pulse duration. Notably, the intensity in the frequency range $\SI{3.0}{\giga\hertz}<f<\SI{3.5}{\giga\hertz}$ even vanishes at the end of the pulse. (d)  $U=\SI{1.2}{\volt}$, similar to (c), with stronger intensity decrease during pulse application. In contrast to (a-c), no condensation after pulse termination is observed.}
	
\end{figure}

Figures 5(a-d) show the BLS intensity as a function of time and frequency for different applied voltages. In particular, Figs. 5(a-c) show the whole pulse duration for the measurements presented in the main text with the applied voltages of $U=\SI{0.9}{\volt}$ [Fig. 5(a), Fig. 2(a) in the main text],  $U=\SI{1.0}{\volt}$ [Fig. 5(b), Fig. 3(a) in the main text] and $U=\SI{1.1}{\volt}$ [Fig. 5(a), Fig. 3(c) in the main text]. For all cases, an increased magnon density in a broad spectral range at the beginning of the applied pulses is observed, analogous to our previous experiments presented in Ref. [28]. After a specific time, which depends on the applied voltage, a distinct peak at a lower frequency appears. We attribute this peak to the excitation of the bullet mode. At the time the bullet is formed, the intensity at higher frequencies decreases. Further, the formation time of the bullet decreases with increasing voltage. The qualitatively same behavior has been observed in similar experiments [29]. 
The spectral magnon density in the frequency range of the magnon gas changes with a variation of the applied voltage. For a low voltage of $\SI{0.9}{\volt}$ [Fig. 5(a)], the intensity increase is the strongest at a frequency of $f\approx\SI{3.5}{\giga\hertz}$, which is below the frequency of thermal magnons of $f\approx\SI{4.0}{\giga\hertz}$. For $\SI{1.0}{\volt}$, the spectral width of the magnon gas related signal at the end of the pulse increases. We observe two poorly pronounced peaks for $f\approx\SI{3.25}{\giga\hertz}$ and $f\approx\SI{3.75}{\giga\hertz}$. These peaks might correspond to certain distinct modes, none of which seems to get predominantly populated. The exact origin of these modes remains speculative in the scope of the presented experiments. Still, they might result from the interplay of the non-linear shifted dispersion below the Pt-layer and the non-shifted dispersion in the adjacent waveguide.
A further increase in voltage to $U=\SI{1.1}{\volt}$ changes the spectrum given by the magnon gas. Now, the previously dominant ($U=\SI{0.9}{\volt}$) or equally strong increased ($U=\SI{1.0}{\volt}$) signal at $f\approx\SI{3.25}{\giga\hertz}$ vanishes during pulse application. Still, we measure a finite magnon intensity at $f\approx\SI{3.75}{\giga\hertz}$ at the end of the pulse. As shown in the main text, this variation of the predominantly excited magnon states directly influences the subsequent redistribution processes, which set in after pulse termination.
For completeness, Fig. 5(d) shows the same data as Figs. 5(a-c) but for an even higher voltage of $U=\SI{1.2}{\volt}$, a case, which has not been addressed in the main text. As it can be seen, the observed changes in frequency and intensity of the SHE-STT-driven system during the pulse continue. However, the overall intensity during the pulse already significantly decreases, which might be due to an increased temperature. Notably, at this voltage, the RC mechanism triggered redistribution of excess magnons is substantially suppressed, and we observe no stabilization of the bullet mode. This decreasing RC effect might be caused by a too high temperature of the magnon gas at the end of the pulse.

\subsection{SHE-STT-driven excitation of the bullet mode and its stabilization via the RC mechanism for a field aligned along the long axis of the waveguide}

The measurements presented in the main text are conducted for a field aligned along the short axis of the strip. We attribute the observed low-frequency signal to the formation of a standing soliton mode, typically referred to as the magnon bullet mode. This interpretation is based on the fact that, in general, the bullet mode features a frequency below the linear magnon spectrum. However, for a field aligned along the short axis, demagnetization fields need to be considered. These lead to edge modes, which are spatially localized at the edges of the waveguide and exhibit frequencies below the fundamental mode. Therefore, we confirm the bullet mode character of the observed signal by means of additional measurements performed for a field aligned along the long axis. In this geometry, no edge modes exist, and therefore, no signal below the fundamental waveguide mode is expected, except the one caused by the bullet mode excitation.
For the measurements with the external field aligned along the long axis of the waveguide, we use structures of a different contact geometry, as depicted in the right panel of Fig.~6(a). In contrast to the structures used for the experiments presented in the main text (left panel), the current direction is now along the short axis of the strip. Hence, this geometry allows for applying an anti-damping like SHE-STT effect with the field aligned along the waveguide. The structure investigated here was fabricated simultaneously and on the same chip as the structure addressed in the main text. The external field aligned along the $y$-axis is $\upmu_0 H_\mathrm{ext}=\SI{162}{\milli\tesla}$, and the pulse duration of the applied DC-pulse is $\tau_\mathrm{P}=\SI{50}{\nano\second}$.

\begin{figure}[]

		\includegraphics[width=\textwidth]{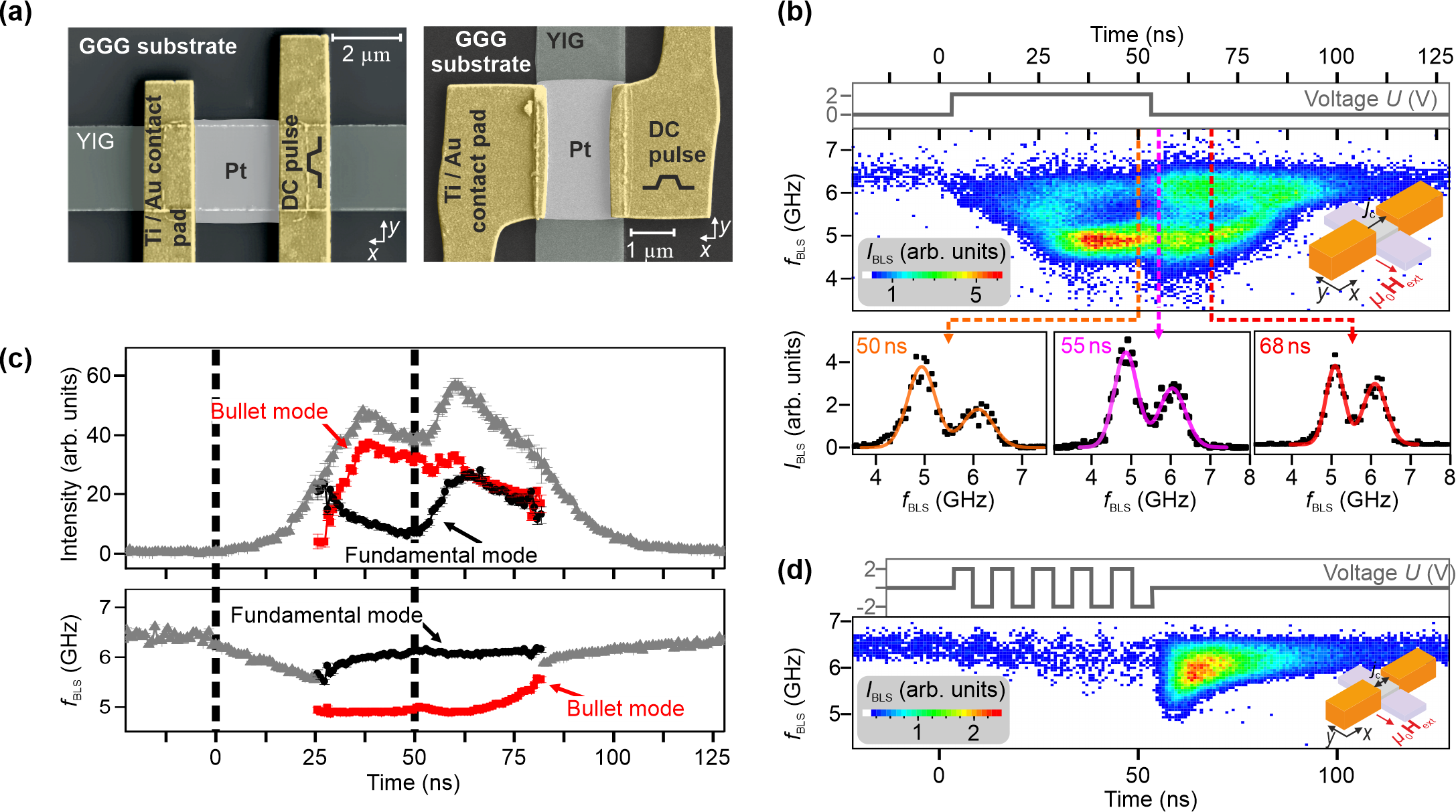}

	\caption{(a) Structure with different contact geometries addressed in the main text (left panel) and used for the control experiment with a field aligned along the long axis (right panel). The current along the short axis of the waveguide allows for a SHE-STT effect triggered magnon injection for the new field direction, with $\upmu_0 H_\mathrm{ext}=\SI{162}{\milli\tesla}$  and $\mathbf{{H}}_\mathrm{ext}$ along the $y$-axis. (b) Middle panel: BLS intensity as a function of frequency and time for a voltage of  U=2.0 V  and pulse duration of $\tau_\mathrm{P}=\SI{50}{\nano\second}$, upper panel: Applied voltage as a function of time, lower panel: exemplarily extracted spectra a different times (black squares) and fits (lines). (c) Extracted intensities (upper panel) and frequencies (lower panel) as a function of time for the bullet mode (red squares) and the fundamental mode (black dots). The gray triangles show the sum of both intensities or the result of a single Gaussian fit at times applicable. (d) Control experiment, analogue to (b), but now with alternating current polarity, effectively suppressing the SHE-STT effect. As a result, no bullet mode excitation is observed.  }
	
\end{figure}

Figure~6(b) shows the BLS intensity as a function of time and frequency. The upper panel shows the applied voltage schematically as a function of time. As one can see, qualitatively the same effect of the SHE-STT mechanism as in the experiments presented in the main text is observed here: The application of the DC-pulse leads first to an increased BLS intensity in a broad frequency range, and consequently, to a shift of the frequency down. Here, at a time of $t\approx\SI{25}{\nano\second}$, the bullet mode is formed, visible as a distinct peak at a frequency of $f\approx\SI{5}{\giga\hertz}$. Because we can exclude any presence of edge-modes for the used geometry, we can confidently claim that this peak is attributed to the excitation of the bullet mode. Therefore, we conclude that the low-frequency signal for the field direction along the short axis is attributed to the bullet mode excitation. After pulse termination, at $t=\SI{50}{\nano\second}$, we observe again a stabilization of this bullet mode, which is triggered via the RC mechanism. 
The lower panel of Fig.~6(b) shows exemplary extracted spectra at the indicated times (black squares) at the end of the applied pulse (left panel), just after its termination (middle panel), and $\SI{18}{\nano\second}$ after the termination (right panel). The lines show the fits of the sum of two Gaussian to the BLS-spectra. The signals corresponding to the bottom of the magnon gas ($f\approx\SI{6}{\giga\hertz}$) and to the bullet mode ($f\approx\SI{5}{\giga\hertz}$) can be well separated over a long time interval, and the bullet is stabilized until $t=\SI{70}{\nano\second}$.
We conduct the exemplarily shown fits at every time step, and the so obtained frequencies and intensities as a function of time are depicted in Fig.~6(c). The black dots and red squares in the upper panel show the fundamental mode and bullet mode intensities, respectively. In addition, the gray triangles are the sum of both intensities or the intensity obtained by fitting a single Gaussian at times applicable. Again, we find the excess magnons generated via the RC mechanism are redistributed to the bottom of the magnon gas and to the bullet. This redistribution can be seen by the increasing intensity of both modes initially after pulse termination and the fact that both modes exhibit a similar decay over time (in contrast to the reduced lifetime of the bullet in the absence of an external injection, which is discussed in the main text).
The lower panel of Fig.~6(c) shows the extracted frequencies of both modes (same color code as in the upper panel). We find again that the bullet frequency initially increases after pulse termination. In the following, the redistribution of the excess magnons stabilizes the bullet, which can be seen from the decreasing bullet frequency after pulse termination until $t=\SI{60}{\nano\second}$.
Finally, we conduct an additional control experiment for the same structure and field direction but with an effectively vanishing SHE-STT effect. For this purpose, we apply a DC-pulse of the same amplitude and duration but with an alternating polarity (10 intervals, each of a period of \SI{5}{\nano\second}). In such a way, the SHE-STT effect alternately injects and annihilates magnons. The measured BLS intensity as a function of time and frequency is shown in Figure~6(d). As it can be seen, the BLS intensity during pulse application is decreased, as we would expect for a purely temperature-induced increase of the magnon density. The slight variation of the frequency over time during the pulse is a consequence of the alternating direction of the generated Oerstedt fields and of the alternating SHE-STT effect. After pulse termination, we observe the generated excess magnons redistribute to the bottom of the fundamental mode. At smaller frequencies we observe no accumulation of magnons, which shows that the bullet mode is absent in this case, when no external injection is present during the pulse.

\subsection{Localization of the bullet mode and the fundamental mode}

Auto-oscillations driven by the STT-effect often result in the excitation of the so-called magnon bullet mode. Due to non-linear effects, this mode features a self-localized behavior. Hence, the interpretation of the bullet mode nature of the low-frequency signal observed in the presented experiments can be validated by measuring the spatial distribution of the excited mode. For that purpose, additional space- and time-resolved BLS measurements were conducted on a similar structure of the same geometry and fabricated simultaneously on the same chip. The external field applied was $\upmu_0 H_\mathrm{ext}=\SI{55}{\milli\tesla}$, slightly lower as in the experiments presented in the main text.

\begin{figure}[]
	
		\includegraphics[width=\textwidth]{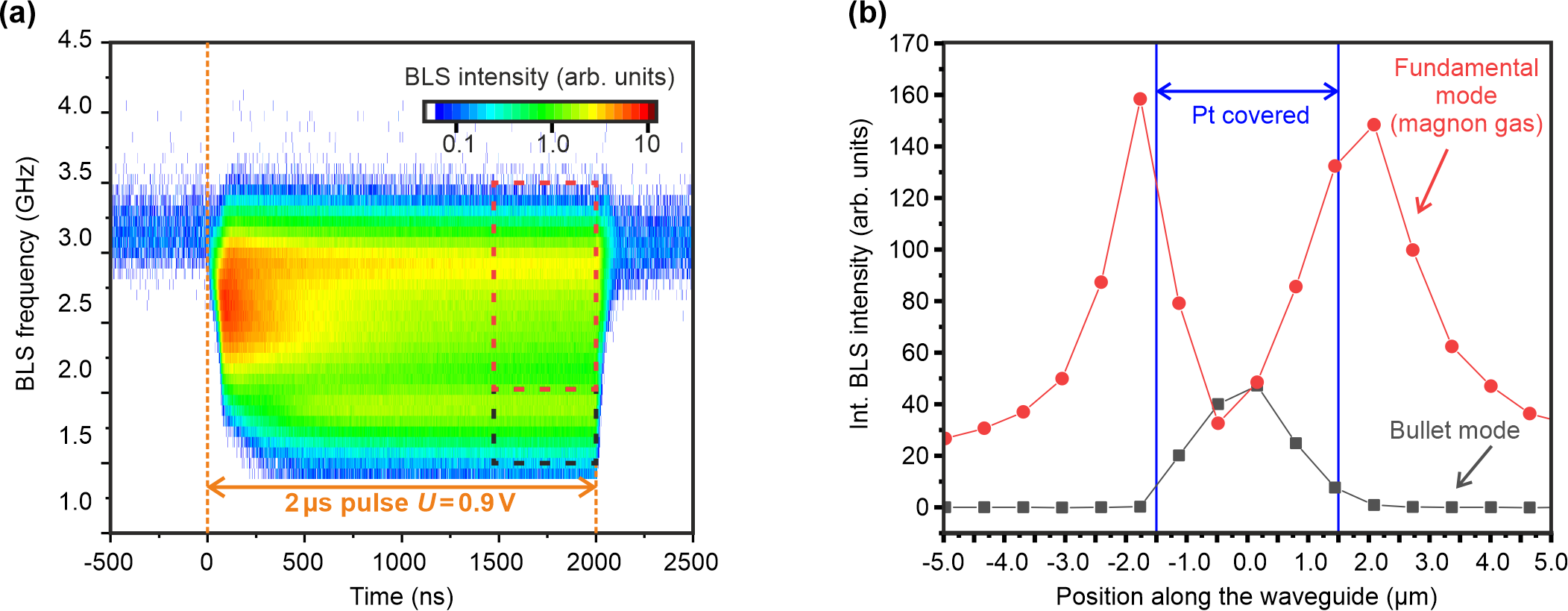}

	\caption{(a) BLS intensity as a function of time and frequency. The orange dashed lines depict the time a 2-$\upmu$s-long DC-pulse of an amplitude of $U=\SI{0.9}{\volt}$  is applied. The dashed rectangles mark the integration time and frequency to determine the bullet mode intensity (black) and the fundamental mode intensity (red). The external field applied was  $\upmu_0 H_\mathrm{ext}=\SI{55}{\milli\tesla}$.  (b) Integrated BLS intensity as a function of the position along the waveguide. The integration ranges for the bullet mode (black squares) and the fundamental mode (red dots) are shown in (a). The vertical blue lines indicate the waveguide region covered with Pt. The bullet mode intensity is localized below the Pt layer, and the fundamental mode intensity extends to the bare YIG waveguide.   }
	
\end{figure}
 Figure~7(a) shows the BLS intensity as a function of time and frequency measured in the center of the Pt-layer. The dashed orange lines indicate the time a 2-$\upmu$s-long DC pulse of an amplitude of $U=\SI{0.9}{\volt}$   is applied (We find the threshold of STT-induced damping compensation for this specific structure and external field as $U_\mathrm{th}=\SI{0.46}{\volt}$). As it can be seen, the spectral distribution of excited magnons matches qualitatively the experiments presented in the main text. In the following, we scan the laser spot along the waveguide. We integrate the BLS intensity in the frequency and time frames indicated by the dashed rectangles. The obtained integrated BLS intensities, shown in black and red, correspond to the STT-driven quasi-stationary bullet mode and fundamental mode, respectively.
Figure 7(b) shows the integrated intensity as a function of the position along the waveguide for the bullet mode (black squares) and the fundamental mode (red dots). The blue horizontal lines indicate the region of the Pt-layer. As it can be seen, the bullet mode is localized at the center of the Pt-layer. Outside the Pt-covered region, the bullet mode intensity vanishes. In contrast, the fundamental mode (magnon gas) intensity is the largest in the bare YIG waveguide, in the vicinity of the Pt-covered region. Within the region of the Pt-injector, the fundamental mode intensity decreases but is still of the same order as the bullet mode intensity. These opposing spatial distributions imply a localization of the bullet mode and that the presence of the bullet decreases the fundamental mode intensity. The latter can be understood in terms of a constant STT effect distributing among both modes in the region where the bullet is present. The fact that the fundamental mode intensity is the largest outside the Pt-injector can be due to mode-specific BLS sensitivity and the temperature-induced decrease of the BLS sensitivity. Considering that the temperature is higher below the Pt-layer, we expect a lower BLS sensitivity in this region.
The spatial resolution of the BLS setup is in the range of $\SI{250}{\nano\meter}$. Therefore we cannot determine the spatial width of the bullet mode more precisely. However, the fact that we observe a localization of the bullet mode and at the same time a spatially extended intensity distribution of the fundamental mode supports our previously made assumption that the low-frequency excitation is attributed to the formation of the magnon bullet mode.

\subsection{References }
[S1] T. Meyer, T. Brächer, F. Heussner, A. A. Serga, H. Naganuma, K. Mukaiyama, M. Oogane, Y. Ando, B. Hillebrands, and P. Pirro, Characterization of spin-transfer torque effect induced magnetization dynamics driven by short current pulses, Appl. Phys. Lett. 112, 22401 (2018).

\end{document}